\documentclass[showpacs]{revtex4}
\usepackage{epsfig}
\def\beq{\begin{equation}}
\def\enq{\end{equation}}
\def\beqa{\begin{eqnarray}}
\def\enqa{\end{eqnarray}}

\def\MeV{\nobreak\,\mbox{MeV}}
\def\GeV{\nobreak\,\mbox{GeV}}

\def\qq{\lag\bar{q}q\rag}

\def\G3{\lag g_s^3G^3\rag}

\def\xsla{x\kern-.5em\slash}
\def\psla{p\kern-.5em\slash}
\newcommand{\rag}{\rangle}
\newcommand{\lag}{\langle}

\begin{document}

\title{
The $\rho(1S,2S)$, $\psi(1S,2S)$, $\Upsilon(1S,2S)$ and 
$\psi_t(1S,2S)$ mesons in a double pole QCD Sum Rule}

\author{M. S. Maior de Sousa}
\affiliation{Unidade Acad\^emica de F\'\i sica, Universidade Federal de Campina Grande,\\
58.051-970, Campina Grande, PB, Brazil}

\author{R. Rodrigues da Silva}
\email{romulo@df.ufcg.edu.br}
\affiliation{Unidade Acad\^emica de F\'\i sica, Universidade Federal de Campina Grande,\\
58.051-970, Campina Grande, PB, Brazil}

\begin{abstract}

We use the method of double pole QCD sum rule which is basically a fit with two exponentials of the correlation function, where we can extract the masses and decay constants of mesons as a function of the Borel mass. 
We apply this method to study the mesons: $\rho(1S,2S)$, $\psi(1S,2S)$, $\Upsilon(1S,2S)$ and $\psi_t(1S,2S)$. 
We also present predictions for the toponiuns masses   
$\psi_t(1S,2S)$ of $m(1S)=357\GeV$ and $m(2S)=374\GeV$.

\end{abstract}

\maketitle

\section{Introduction}

In 1977, Shifman, Vainshtein, Zakharov, Novikov, Okun  and Voloshin \cite{Novikov:1977dq,Novikov:1977cm,Shifman:1978bx,Shifman:1978zq} 
created the successful method of QCD sum rules (QCDSR), which is widely used nowadays. With this method, we can calculate many hadron parameters such as: mass of the hadron,  decay constant, coupling constant and form factors in terms of the QCD parameters as for example: quark masses, the strong coupling and non-perturbative parameters like quark condensate and gluon condensate. The main point of this method is that the quantum numbers and content of quarks in hadron are represented 
by an interpolating current, 
where the correlation function of this current
is introduced in the framework of the operator product expansion (OPE). 
To determine the mass and the decay constant of the ground state of the hadron, we use the two point correlation function. On the QCD side, the correlation function can be written in terms of a dispersion relation and on the phenomenological side can be written 
in terms of the ground state and several excited states.
The usual QCDSR method uses an ansatz that the phenomenological spectral density can be represented by a form pole plus continuum, where is assumed that the phenomenological and QCD spectral density coincides with 
each other above the continuum threshold.
The continuum is represented 
by an extra parameter called, $s_0$,
as being correlated with the onset of 
excited states \cite{Colangelo:2000dp}.
In general, the resonance activity occurs 
with $\sqrt{s_0}$ 
lower than the mass of the first excited state. 

\begin{figure}[!htb]
\begin{minipage}[b]{7cm}
\includegraphics[height=4.5cm]{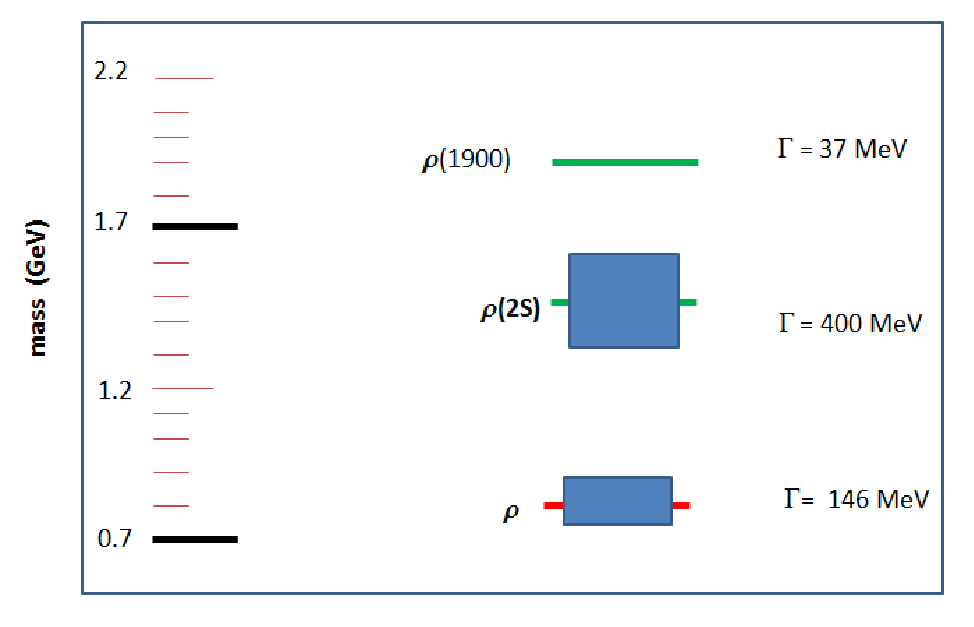}
\end{minipage}
\begin{minipage}[b]{7cm}
\includegraphics[height=4.5cm]{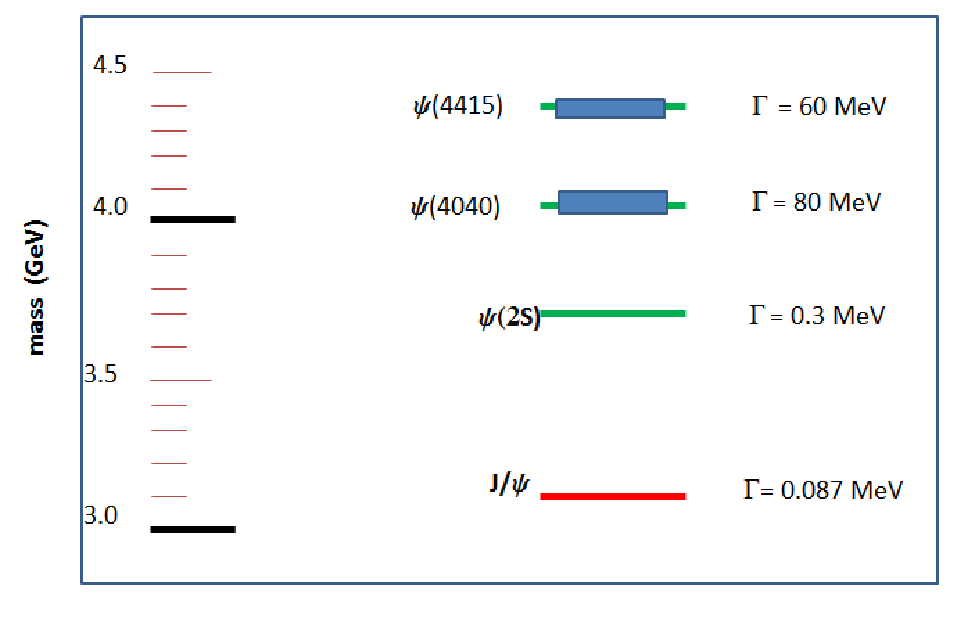}
\end{minipage}
\label{spec}
\caption{\small (Left) The radial excited states of the $\rho$ meson \cite{pdg.2012,Godfrey:1985xj}. The $\rho(1540)$ and $\rho(1900)$ are omitted from the PDG summary table, but $\rho(1900)$ is a good candidate for $\rho(3S)$ was predicted by the Refs. \cite{Godfrey:1985xj,Ebert:2009ub}. For $\rho(1540)$, its existence is not predicted by usual $q\bar{q}$ model. (Right) The radial excited states of the $J/\psi$ meson \cite{pdg.2012,Godfrey:1985xj,Lakhina:2006vg}.}
\end{figure}

For the $\rho$ meson spectrum, Fig.(\ref{spec}), the purpose
of pole plus continuum is a good approach, due to the large decay width of the $\rho(2S)$ or $\rho(1450)$, 
that allow to approximate the excited states as a continuum.

For the $\rho$ meson \cite{Reinders:1984sr}
the value of $\sqrt{s_0}$ 
that best fit the mass and the decay constant 
is $\sqrt{s_0}=1.2$ GeV 
and for the $\phi(1020)$ meson  
the value is $\sqrt{s_0}=1.41$. 
We note that the values quoted above for $\sqrt{s_0}$ 
are of about 250 MeV below 
of the poles of $\rho(1450)$ and $\phi(1680)$. 
One interpretation of this result, 
it is due to the effect of the 
large decay width of these mesons.

The pioneering work on charmonium sum rule, Novikov et al. \cite{Novikov:1977dq} 
considered the phenomenological 
side with double pole and $\sqrt{s_0'}=4$ GeV, 
where $\sqrt{s_0'}$ is double pole continuum 
parameter. This value is correlated with the 
threshold of pair production of charmed mesons.
Using this value of $\sqrt{s_0'}$ and Moment Sum Rule
at $Q^2=0$, they presented the first estimate 
for the gluon condensate
and a nice prediction of the 
$\eta_c$ mass of 3.0 GeV, 
while the experimental results in 1977 reported 
a $\eta_c$ mass of 2.83 GeV
\cite{Novikov:1977dq,Novikov:1977cm,Shifman:1978zq}.

In single pole sum rule for $J/\psi$ and 
$\eta_c$, the best values 
of $\sqrt{s_0}$ that fit the masses are 
$(3.8 \pm 0.2)$ GeV \cite{Reinders:1984sr}, 
where the minimum value of $\sqrt{s_0}$ 
is 100 MeV below of the $\psi(2S)$ mass. 
As the decay width of the $\psi(2S)$ is about 0.3 MeV, 
so it is approximate to associate the parameter 
$\sqrt{s_0}$ with some activity of excited states.

In principle the value of $\sqrt{s_0}$ 
can be fixed by setting the mass of the ground state,
on the other hand, in the case where the mass of 
the ground state is unknown 
as in the case of tetraquarks, there are studies 
that extract the lower limit of $\sqrt{s_0}$,
since the pole dominance and OPE convergence 
is controlled \cite{Bracco:2008jj}. 

In double pole QCDSR \cite{Shifman:1978zq}, 
we expect that a reliable sum rule should provide 
that the ground state decay constant is larger 
than the excited state decay constant and
provides an upper limit of $\sqrt{s_0'}$, 
as we can see in our results.
This condition is directly related 
with the expression of decay constant 
obtained from potential models \cite{Zakharov:1975ku,Segovia:2014mca,Lakhina:2006vg}
is proportional to the meson wave function at the origin. 
As the meson radius increases with excitation, 
the probability of finding its quarks 
at the origin declines with the 
excitation \cite{Leinweber:1994gt}.
This condition was used by Shifman \cite{Shifman:1978zq}
to predict the mass of the $\eta_c$ 
and this condition agrees with the 
experimental data from the spectrum of 
$\psi(nS)$ and $\Upsilon(nS)$
up to $\Upsilon(4S)$ \cite{pdg.2012}. 
For $\Upsilon(5S)$ 
is observed a violation in this behavior, 
where the decay constant 
of $\Upsilon(5S)$ is larger than
$\Upsilon(4S)$. This result is not predicted by potential
models and the authors of Ref.\cite{Segovia:2014mca} 
suggest that the $\Upsilon(5S)$ could be a tetraquark 
or molecule state. 

%
In QCDSR, the excited states are studied in: 
pole-pole plus continuum in Moment Sum Rule 
at $Q^2=0$ \cite{Novikov:1977dq,Novikov:1977cm},
the spectral sum rules with 
pole-pole-pole plus continuum \cite{Singh:2006ii},
the Maximum Entropy Method \cite{Gubler:2010cf} and 
Gaussian Sum Rule with pole-pole plus continuum ansatz \cite{Harnett:2008cw}. There are studies on the 
$\rho(1S,2S)$ mesons \cite{Gubler:2010cf,Bakulev:1998pf,Pimikov:2013usa}, nucleons \cite{Singh:2006ii,Ohtani:2011yy}, 
$\eta_c(1S,2S)$ mesons \cite{Novikov:1977cm}, $\psi(1S,2S)$ mesons \cite{Novikov:1977dq,Gubler:2011ua} and 
$\Upsilon(1S,2S)$ mesons \cite{Suzuki:2012ze}.
In Gaussian Sum Rule is studied the mixed states of the glueballs and scalar mesons. In lattice QCD, there are studies on the $\pi(1S,2S)$ mesons \cite{McNeile:2006qy}, $\rho$ mesons excited states \cite{Burch:2004he,Yamazaki:2001er}, charmonium \cite{Dudek:2006ej,Dudek:2007wv,Liu:2011rn}, nucleons excited states \cite{Mathur:2003zf,Edwards:2011jj,Leinweber:1994gt,Guadagnoli:2004wm} and exotic charmonium spectrum \cite{Liu:2012ze}. In addition, the excited states have been studied recently by several approaches like: QCD Bethe-Salpeter equation \cite{Qin:2011xq} for $\pi(2S)$ and $\rho(2S)$, light-front quark model \cite{Peng:2012tr,Arndt:1999wx} for $\rho(2S)$, $\eta_{c}(2S)$, $\psi(2S)$ and the bottomonium analogous. The $\psi(2S)$ has been studied in QCDSR as a hybrid meson \cite{Kisslinger:2009pw} using the pole plus continuum ansatz.

The method pole-pole plus continuum ansatz was used in lattice QCD
for nucleons \cite{Leinweber:1994gt}. The authors have shown a problem in 
which the ground state coupling strength is lower than the excited 
state coupling strength. 

There are many motivations to study the excited states that belong the charmonium spectrum.  New charmonium-like states Y(4260) and Y(4660) are an example of the importance of excited states. When considering theories that Y(4260) has been proposed as a bound state of $J/\psi-f0$ \cite{MartinezTorres:2009xb} and Y(4660) has been interpreted as a bound state of $J/\psi(2S)-f0$, \cite{Wang:2009hi,Albuquerque:2011ix,Guo:2008zg}, where we can speculate that Y(4660) is an excited state of Y(4260). Another point is that $Z^{+}(4430)$ could be an excited state of $Z_c(3900)$ and $Z_b^{+}(10610)$ could be an excited state of $X_b^{+}(10100)$ \cite{Navarra:2011xa}.

In this paper, we study the excited state using the pole-pole plus continuum 
ansatz in QCD sum rules and we apply in four cases: 
the $\rho(1S,2S)$, $\psi(1S,2S)$, $\Upsilon(1S,2S)$
and $\psi_t(1S,2S)$ mesons and we calculate 
their masses and decay constants.   

\section{The Sum Rule}

In the determination of the mass and the decay constant with QCDSR, we use the two point correlation function \cite{Shifman:1978bx},

\beq
\Pi_{\mu\nu}(q)= i\int d^4 x \; e^{iq\cdot x} 
\langle 0 \!\mid T\{j_{\mu}(x)j^{\dagger}_{\nu}(0)\}\mid \! 0\rangle, 
\label{Pi_c}
\enq

where on the QCD point of view, the current of $q\bar{q} $ vector mesons has the form:
 
\beq
j_\mu(x) =  \bar{q}_a (x)\:\gamma_\mu \:q_a(x)
\label{field}
\enq

Inserting this current in the correlation function, Eq.(\ref{Pi_c}), are obtained the operators expansion, OPE, which can be written in terms of a dispersion relation which depends on the QCD parameters,  then the correlator can be written in the form:

\beq
\Pi_{\mu\nu}^{QCD}(q)= i\int d^4 x \; e^{iq\cdot x} 
\langle 0 \!\mid T\{j_{\mu}(x)j^{\dagger}_{\nu}(0)\}\mid \! 0\rangle
= (q_\mu q_\nu -q^2\,g_{\mu \nu})\Pi^{QCD}(q^2)\,, 
\label{Piqcd}
\enq

with:

\beq
\Pi^{QCD}(q^2)=\int\limits_{s_0^{min}}^\infty ds\, 
\frac{\rho^{Pert}(s)}{s-q^2}
+
\Pi^{nonPert}(q^2), 
\label{Piqcd2}
\enq
where
$\rho^{Pert}(s)=\frac{\mbox{Im}(\Pi^{Pert}(s))}{\pi}$ and the parameter $s_0^{min}$ 
is the minimum value of $s$ to have an imaginary part of the perturbative term $\Pi^{Pert}(s)$ and the correlator $\Pi^{nonPert}(q^2)$ 
is the contribution of the condensates.

On the phenomenological side, we use:

\beq 
\langle 0 \! \mid  j_{\mu}(0) \mid \! V(q) \rangle =  
f_{V} m_{V}\epsilon^{(V)}_\mu(q),
\label{matrix}
\enq 

with $f_{V}$ is the decay constant and $m_{V}$ is the meson mass. Inserting Eq.(\ref{matrix}) in Eq.(\ref{Pi_c}), we get:
 
\beq
\Pi_{\mu\nu}^{Phen}(q)=
(q_\mu q_\nu- q^2\,g_{\mu \nu})
\frac{f_{V}^2}{m_{V}^2 -q^2}
+\mbox{excited states contribution.}
\label{Pifencc}
\enq

We can write the invariant part of the correlator of the Eq.(\ref{Pifencc}) in the form:

\beq
\Pi_{\mu\nu}^{Phen}(q)=
(q_\mu q_\nu- q^2\,g_{\mu \nu})
\Pi^{Phen}(q^2),
\label{Pifen2}
\enq
with
\beq
\Pi^{Phen}(q^2)= 
\int\limits_{0}^\infty ds\, 
\frac{\rho^{Phen}(s)}{s-q^2}, 
\label{Pifen3}
\enq

and $\rho^{Phen}(s)=f_{V}^2\,\delta(s-m_{V}^2) + \rho^{Excited}(s)$. 

When comparing the Eq.(\ref{Piqcd}) with Eq.(\ref{Pifen2})    
the simplest way to perform the sum rule is choosing an invariant structure and equating both sides of the sum rule,
so we have: 

\beq
\Pi^{Phen}(q^2)
=
\Pi^{QCD}(q^2)
\enq

Finally, we obtain the sum rule:

\beq 
\int\limits_{0}^\infty ds\, 
\frac{\rho^{Phen}(s)}{s-q^2}
= 
\int\limits_{s_0^{min}}^\infty ds\, 
\frac{\rho^{Pert}(s)}{s-q^2}
+
\Pi^{nonPert}(q^2).
\label{SR}
\enq

To improve the equivalence between the two sides of 
the sum rule is convenient to use the Borel transformation \cite{Shifman:1978bx}:

\beq 
\label{BSR}
\int\limits_{0}^\infty ds\, 
\rho^{Phen}(s)e^{-s\tau}
= 
\int\limits_{s_0^{min}}^\infty ds\, 
\rho^{Pert}(s)e^{-s\tau}
+
\Pi^{nonPert}(\tau),
\enq

with $\Pi^{nonPert}(\tau)=B[\Pi^{nonPert}(q^2)]$ and $\tau=1/M^2$, where M is Borel mass.
\\ 

For the sum rule of $\rho$ meson we use 
$\rho^{Pert}(s)$ and $\Pi^{nonPert}(\tau)$ are given by \cite{Shifman:1978bx,Reinders:1984sr,Colangelo:2000dp}:

\beq
\label{pert_rho}
\rho^{Pert}(s)
=\frac{1}{4\pi^2}\left(1+ \frac{\alpha_s}{\pi}\right),
\enq

\beq
\label{nonpert_rho}
\Pi^{nonPert}(\tau)=   
\tau
\left(
\frac{1}{12}\langle \frac{\alpha_s}{\pi} G^2\rangle 
+ 2\,m_q\langle \bar{q}q\rangle
\right)
-\tau^2\frac{112}{81}\pi\alpha_s
\langle \bar{q}q\rangle^2,
\enq

where $\alpha_s$ is the strong coupling constant, $m_q$ is light quark mass, 
$\langle \frac{\alpha_s}{\pi} G^2\rangle$ is gluon condensate,
$\langle \bar{q}q\rangle$ is quark condensate,
and $s_0^{min}=4\,m_q^2$. 
We use these parameters
at $\mu=1$ GeV renormalization scale 
\cite{Colangelo:2000dp}.

For the sum rules of $J/\psi$ and $\Upsilon$ mesons we use 
$\rho^{Pert}(s)$ and $\Pi^{nonPert}(\tau)$
are given by \cite{Reinders:1984sr,Novikov:1977dq}.

\beq
\label{pert_psi}
\rho^{Pert}(s) =
\rho_{0}(s)+ \rho_{rad}(s),
\enq

where,

\beq
\rho_{0}(s)=\frac{1}{8\pi^2} v(3-v^2),
\enq

\beq
\label{Reinders}
\rho_{rad}(s)=
\frac{4\alpha_s}{3}\rho_{0}(s)
\left[
\frac{\pi}{2\,v} -\frac{3+v}{4}
\left(
\frac{\pi}{2} -\frac{3}{4\pi}
\right)
\right]
-\frac{3\alpha_s}{4\pi^3}
\frac{(1-v^2)^2}{v}
\mbox{ln}(2).
\enq

with $v=\sqrt{1- 4m^2/s}$ and m is off-shell heavy quark mass and $\alpha_s=\alpha_s(m)$.

For the gluon condensate we apply the Borel transform 
of the expression was given by
Reinders et al. \cite{Reinders:1984sr}, 
where we have:

\beq
\label{nonpert_psi}
\Pi^{nonPert}(\tau)=   
-\frac{\tau}{12}
\langle \frac{\alpha_s}{\pi} G^2\rangle 
\int_{0}^{1} d\alpha
\left(
1+ \frac{m^2\,\tau}{\alpha(1-\alpha)}
\right)
\exp{\left(\frac{-m^2\,\tau}{\alpha(1-\alpha)}\right)}
. 
\enq

\section{The Method}

To implement our method, we consider the following spectral density on the phenomenological side:   

\beq
\label{rho_phen}
\rho^{Phen}(s)=\lambda_{1}^2 \delta(s-m_{1}^2)+
\lambda_{2}^2 \delta(s-m_{2}^2)+
\rho^{Cont}(s)\theta(s-s'_0),
\enq

where $m_{1}$ is the mass of the ground state and $m_{2}$ is the mass of the first excited state, 
$\lambda_{1}$ coupling strength is the decay constant for the ground state and $\lambda_{2}$ is the decay constant for the first excited state and $s_0'$ 
mark the onset of the continuum states.
Inserting Eq.(\ref{rho_phen}) on the left hand side of Eq.(\ref{BSR}),  we get the expressions:

\beq
\label{phen}
\Pi^{LHS}(\tau)=
\lambda_{1}^2 e^{-m^2_{1}\tau}
+ 
\lambda_{2}^2 e^{-m^2_{2}\tau}    
+
\int\limits_{s_0'}^{\infty} ds\, 
\rho^{Phen}(s)\,e^{-s\tau},
\enq

On the right hand side of 
Eq.(\ref{BSR}), we get:  

\beq
\label{pert}
\Pi^{RHS}(\tau)= 
\int\limits_{s_0^{min}}^{s_0'} ds\,\rho^{Pert}(s)e^{-s\tau}+
\int\limits_{s_0'}^{\infty} ds\,\rho^{Pert}(s)e^{-s\tau}+
\Pi^{nonPert}(\tau).
\enq

Equating Eq.(\ref{phen}) with Eq.(\ref{pert}) and using the quark hadron duality, 
where we assume that 
$\rho^{Phen}(s)=\rho^{Pert}(s)$ 
for $s\geq s_0'$,
so we get the double pole QCD sum rule,

\beq
\label{ppSR}
\lambda_{1}^2 \, e^{-m^2_{1}\tau}+ 
\lambda_{2}^2 \, e^{-m^2_{2}\tau}
=
\Pi^{QCD}(\tau),
\enq

where,

\beq
\Pi^{QCD}(\tau)=
\int\limits_{s_0^{min}}^{s_0'}ds~
  e^{-s\tau}\rho^{Pert}(s)
+
\Pi^{nonPert}(\tau)
.
\label{corr2}
\enq

The contribution of the resonances is given by:

\beq
CE(\tau)=
\int\limits_{s_0'}^{\infty} ds\, 
\rho^{Pert}(s)\,e^{-s\tau}.
\label{reso}
\enq

As usually done in QCDSR, the obtaining mass of the hadron, we take the derivative of Eq.(\ref{ppSR}) with respect to $\tau$ and we get the new equation:

\beq
\label{ppSR2}
-m^2_{1}\lambda_{1}^2 e^{-m^2_{1}\tau}
-m^2_{2}\lambda_{2}^2 e^{-m^2_{2}\tau}
=
\frac{d}{d\tau}\Pi^{QCD}(\tau).
\enq

We can observe that the equations Eq.(\ref{ppSR}) and Eq.(\ref{ppSR2}) can form an equation  system  in the variables,  

\beq
\label{ppA}
A(\tau)= \lambda_{1}^2 e^{-m^2_{1}\tau},
\enq

\beq
\label{ppB}
B(\tau)= \lambda_{2}^2 e^{-m^2_{2}\tau}.
\enq

Solving the equation system Eq.(\ref{ppSR}) and Eq.(\ref{ppSR2}) writings 
in terms of the functions $A(\tau)$ and $B(\tau)$, we easily get:

\beq
\label{solA}
A(\tau)=\frac{
D\Pi^{QCD}(\tau) + \Pi^{QCD}(\tau)\,m_{2}^2 
}
{
m_{2}^2 -m_{1}^2
}
,
\enq

\beq
\label{solB}
B(\tau)=
\frac{
D\Pi^{QCD}(\tau) + \Pi^{QCD}(\tau)\,m_{1}^2 
}
{
m_{1}^2 -m_{2}^2
},
\enq

where we use the notation
\beq
D^{n}F(\tau)=\frac{d^{n}}{d\tau^{n}}F(\tau).
\enq

To eliminate the dependence of the $\lambda_1$ coupling in Eq.(\ref{solA}), 
we take a derivative of this equation with respect of $\tau$ and divide the result by Eq.(\ref{solA}). 
The result of this procedure is given by the Eq.(\ref{massA}).
The procedure to eliminate $\lambda_2$ coupling is analogous that used above and the result is given by the Eq.(\ref{massB}). 

\beq
\label{massA}
m_{1}
=
\sqrt{
-\frac{
D\Pi^{QCD}(\tau)\,m_{2}^2 +D^2\Pi^{QCD}(\tau) 
}
{
D\Pi^{QCD}(\tau) + \Pi^{QCD}(\tau)\,m_{2}^2
}
}
,
\enq

\beq
\label{massB}
m_{2}=
\sqrt{
-\frac{
D\Pi^{QCD}(\tau)\,m_{1}^2+ D^2\Pi^{QCD}(\tau) 
}
{
D\Pi^{QCD}(\tau)+ \Pi^{QCD}(\tau)\,m_{1}^2
}
}.
\enq

In the first view the Eq.(\ref{massA}) and Eq.(\ref{massB}) suggest a system for the masses $m_1$ and $m_2$, that could be extracted the masses. On the other hand, using Eq.(\ref{massA}) to obtain a $m_{2}$ expression, it reproduces the same result has given in Eq.(\ref{massB}). 
To solve this problem of cannot decouple the masses $m_1$ and $m_2$, we will take the derivative of equation Eq.(\ref{solB}) twice in the form:

\beq
\label{DDsolB}
m^4_{2}\lambda_{2}^2 e^{-m^2_{2}\tau}=
\frac{
D^3\Pi^{QCD}(\tau) + D^2\Pi^{QCD}(\tau)\,m_{1}^2 
}
{
m_{1}^2 -m_{2}^2
},
\enq

Dividing by equation Eq.(\ref{solB}) we have a new mass formula,  is given by:

\beq
\label{DDsolB}
m^4_{2}=
\frac{
D^3\Pi^{QCD}(\tau) + D^2\Pi^{QCD}(\tau)\,m_{1}^2 
}
{
D\Pi^{QCD}(\tau) + \Pi^{QCD}(\tau)\,m_{1}^2
},
\enq

Inserting Eq.(\ref{massA}) in equation Eq.(\ref{DDsolB}) we obtain a polynomial equation with respect to the $m_2$:

\beqa
&&
m^4_{2}\alpha
+
m^2_{2}\beta 
+
\gamma
=
0
,
\label{m2_equation}
\enqa

where 
$\alpha=-D\Pi^{QCD}(\tau)^2
+ \Pi^{QCD}(\tau)\,D^2\Pi^{QCD}(\tau)
$,
$\beta=-D^2\Pi^{QCD}(\tau)\,D\Pi^{QCD}(\tau)
+D^3\Pi^{QCD}(\tau)\,\Pi^{QCD}(\tau)
$,
$\gamma=D^3\Pi^{QCD}(\tau)\,D\Pi^{QCD}(\tau)
-D^2\Pi^{QCD}(\tau)^2
$ and 
$\Delta=\beta^2 -4\alpha\gamma$.

For obtaining $m_1$, we can do the same procedure as above and $m_1$ obeys the same equation Eq.(\ref{m2_equation}). Thus, we easily solved this equation in which the mass of the ground state and the excited state are given by:

\beqa
m_{1}= \sqrt{\frac{-\beta -\sqrt{\Delta}}{2\alpha}},
\label{mass1}
\enqa
\beqa
m_{2}= \sqrt{\frac{-\beta +\sqrt{\Delta}}{2\alpha}}.
\label{mass2}
\enqa

\section{Results}

In this work we use the following parameters 
for $\rho$ meson:
$\alpha_{s}(1\GeV)=0.5$, 
$m_q= (6.4 \pm 1.25)\MeV$ 
$\qq= -(0.240 \pm 0.010)^3\GeV^3$,
$\langle \frac{\alpha_s}{\pi} G^2\rangle=(0.012 \pm 0.004)\GeV^4$
at $\mu=1$ GeV renormalization scale 
\cite{Colangelo:2000dp}.
For $J/\psi$, we use the  
$\alpha_{s}(m_c)=0.3$,
$m_c(mc)=1.3\GeV$
and for $\Upsilon$, we use the
$\alpha_{s}(m_b)=0.15$,  
$m_b(mb)=4.3\GeV$.

In addition to the above mentioned parameters, the sum rule depends 
of the others two parameters: the continuum threshold $s_0'$ 
and the Borel mass, M.

As explained in the introduction, 
we expect that $\sqrt{s_0'}$ is a value closes to
(3S) meson mass, however, in cases where
the 3S state is unknown or has large decay width, the value
of $\sqrt{s_0'}$ is limited by the condition that the decay constant of
2S meson should be smaller than 1S meson and the lowest
limit of $\sqrt{s_0'}$ is considered as m(2S) + 100 MeV.

Using a value of $\sqrt{s_0'}$, the range of Borel mass is chosen on the assumption that the ratio of the double pole Eq.(\ref{ppSR}) and the total contribution pole-pole plus the  resonances, Eq.(\ref{reso}), should be higher than $40\%$.

\subsection{$\rho(1S,2S)$ Sum Rule}

Using the mass of $\rho(3S)$ meson of $1.9 \GeV$, Fig.(\ref{spec}),
we test $\sqrt{s_0'}=1.9\GeV$, but in this case 
the decay constant of excited state is 
larger than the of the ground state,
so the sum rule fails.

The maximum value of $\sqrt{s_0'}$ is 1.66 GeV, where in this
case the decay constant of excited state is slightly below of
the decay constant of ground state. The minimum value of
$\sqrt{s_0'}$ is 1.56 GeV, because
$\sqrt{s_0'}$ - m(2S) reaches the value of
100 MeV.
 
\begin{figure}[!htb]
\begin{center}
\includegraphics[height=5cm]{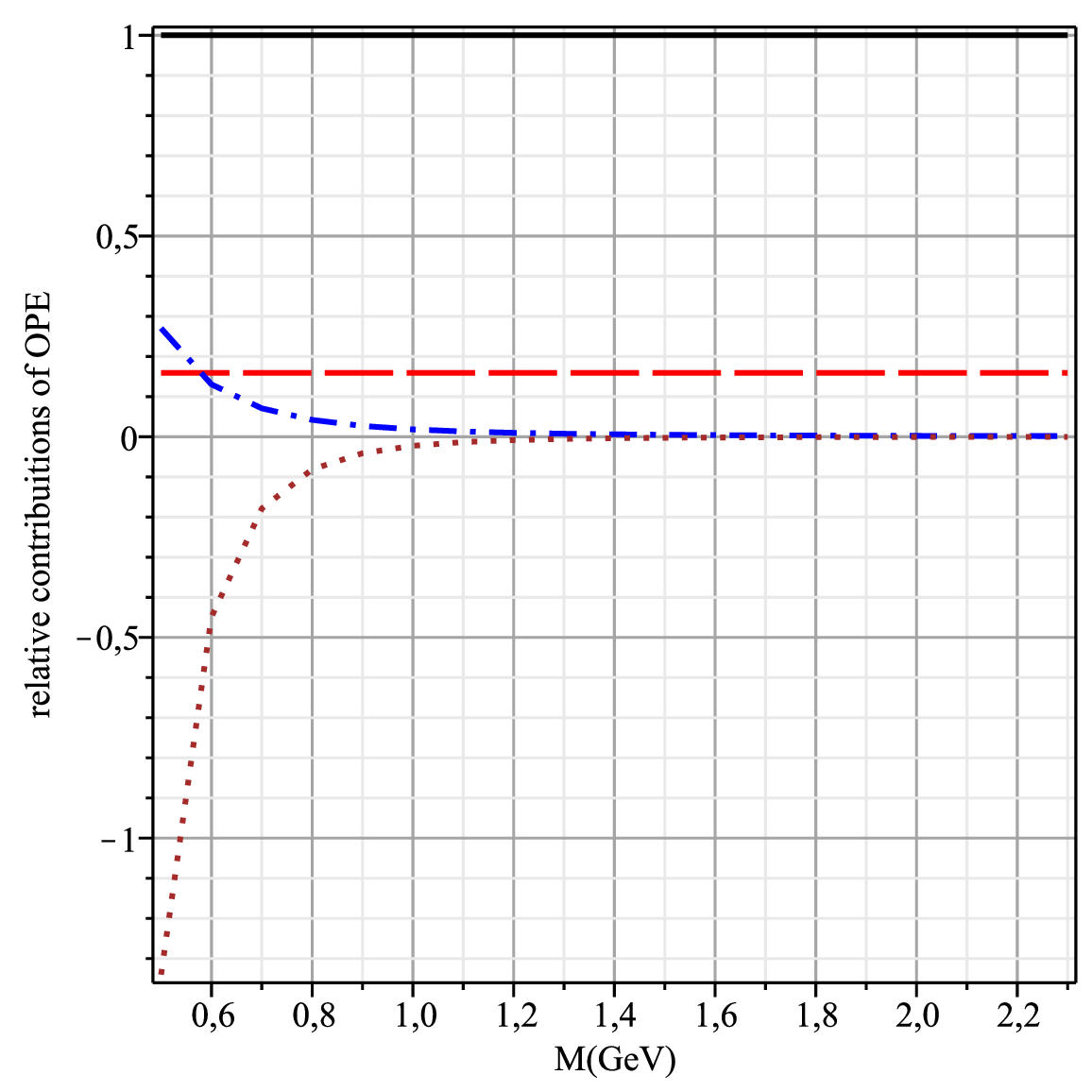}
\caption{\small The relative contributions of OPE for $\rho(1S,2S)$ 
as a function of the Borel mass for $\sqrt{s_0'}=1.61\GeV$.
The solid line for the first order perturbation term,
long-dashed line for radiative correction,
dashed-dot line for dimension 4 and
dot line for dimension 6
.}
\label{OPErho}
\end{center}
\end{figure}

In Fig.(\ref{OPErho}) the contribution of the OPE terms 
are ordered relative to the first order perturbative term of 
Eq.(\ref{pert_rho}) in Eq.(\ref{corr2}). 
The solid line is the contribution of 
the first order perturbation term is adopted as 1,
long-dashed line is the radiative correction,
dashed-dot line is the dimension 4 of Eq.(\ref{nonpert_rho}) and
dot is the dimension 6 of Eq.(\ref{nonpert_rho}).
We note that the convergence of OPE is controlled and 
at M=1 GeV, the contributions of 
the dimension 4 is 1.82\%
and dimension 6 is 2.26\% of the first order perturbation term. 
For M=2 GeV, these condensates
contribute with 300 MeV in the mass of
$\rho(1S)$ and 100 MeV for the mass of $\rho(2S)$.

\begin{figure}[!htb]
\begin{center}
\includegraphics[height=5cm]{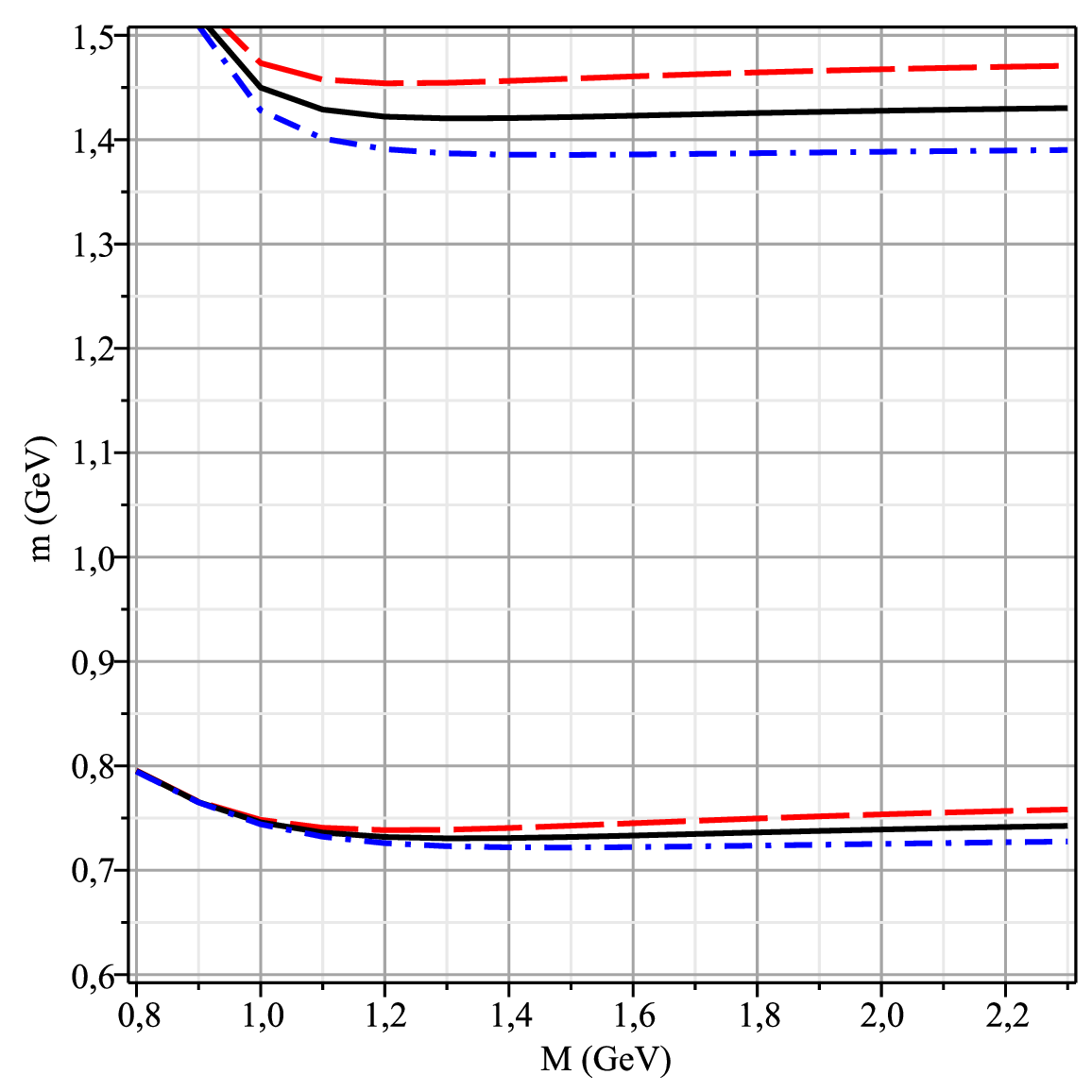}
\caption{\small The $\rho(1S)$, lower lines, and $\rho(2S)$, upper lines, masses as a function of the Borel mass. 
The solid line for $\sqrt{s_0'}=1.61\GeV$, dashed-dot line for $\sqrt{s_0'}=1.56 \GeV$ and long-dashed line for $\sqrt{s_0'}=1.66 \GeV$.}
\label{massRho}
\end{center}
\end{figure}

We study the behavior of the masses and decay constants of the mesons $\rho$ and $\rho(2S)$ as a function of Borel mass for three values $\sqrt{s_0'}$: solid line for $\sqrt{s_0'}=1.61\GeV$, dashed-dot line for $\sqrt{s_0'}=1.56 \GeV$ and long-dashed line for $\sqrt{s_0'}=1.66 \GeV$.
We can see in Fig.(\ref{massRho}) that all masses are stable and at M=1.2 GeV, the long-dashed line gives 
a value compatible with the experimental value for the $\rho(2S)$ mass of 1454 MeV and 
for the $\rho(1S)$ the long-dashed line gives
a mass of 740 MeV.
 
\begin{figure}[!htb]
\begin{center}
\includegraphics[height=5cm]{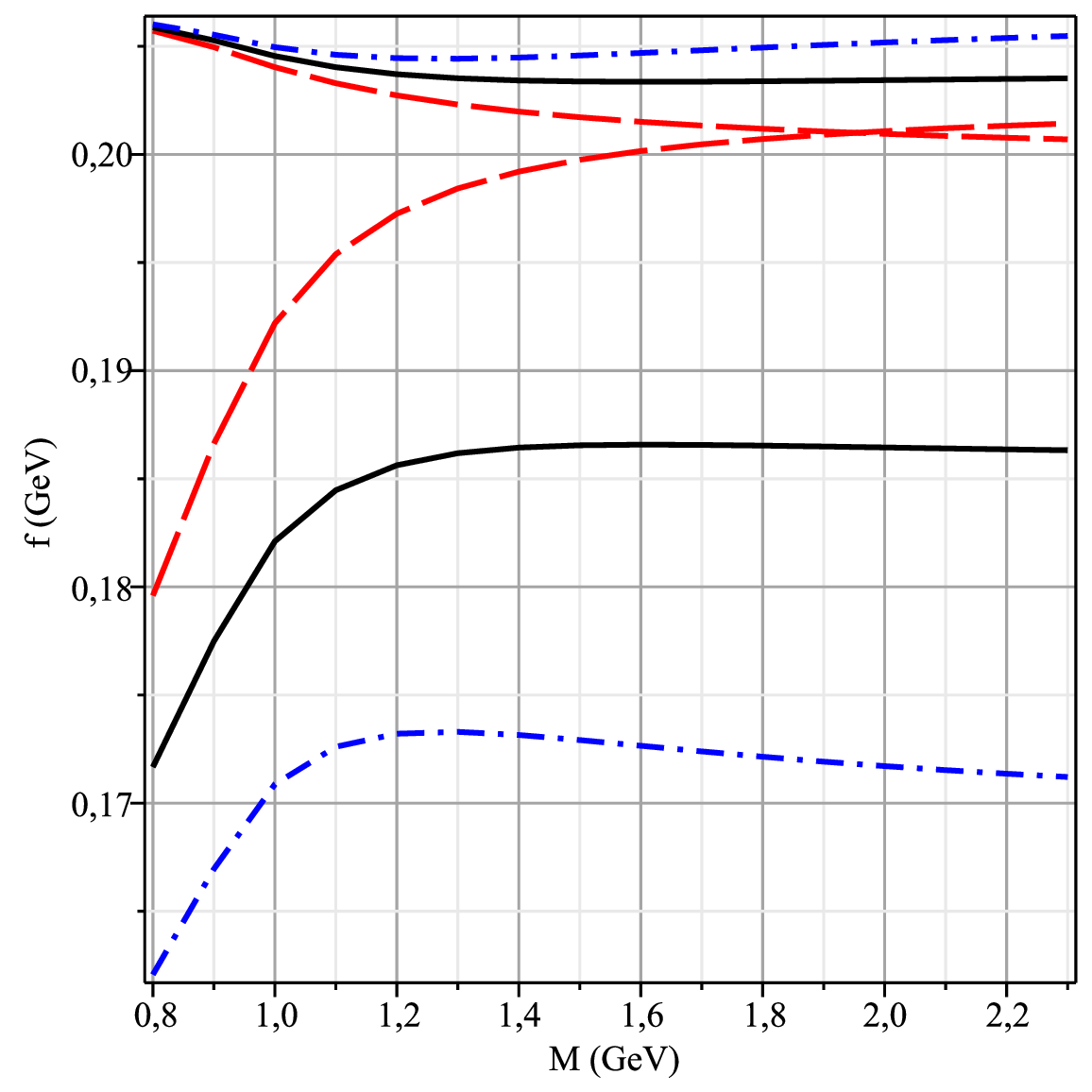}
\caption{\small The $\rho(1S)$, upper lines, and $\rho(2S)$, lower lines, decay constant  as a function of the Borel mass. The solid line for $\sqrt{s_0'}=1.61\GeV$, dashed-dot line for $\sqrt{s_0'}=1.56 \GeV$ and long-dashed line for $\sqrt{s_0'}=1.66 \GeV$.}
\label{decayRho}
\end{center}
\end{figure}

For the calculation of the decay constant, 
we use the experimental values $m_{1}=0.77$ GeV
and $m_{2}=1.46$ GeV. 
In Fig.(\ref{decayRho}), we show the decay constant of the $\rho$ and $\rho(2S)$ mesons. 
Considering the value for $\sqrt{s_0'}$ of 1.61 GeV (solid line), the value of the $\rho$ meson decay constant has a plateau on value 203 MeV and $\rho(2S)$ has a plateau on the value 186 MeV. Considering uncertainty with respect to $\sqrt{s_0'}$ parameter at M= 2 GeV, we get: 

\beq
\label{frho}
f_{\rho}
=
(203 \pm 2) \MeV,
\enq

\beq
\label{frho2S}
f_{\rho(2S)}
=
(186 \pm 14) \MeV.
\enq

The value of $f_{\rho}$ is 17 MeV 
lower than the experimental value 
of $(220.5 \pm 1)$MeV \cite{pdg.2012} 
obtained from $\rho^0\rightarrow e^{+}e^{-}$ 
decay width considering $1/\alpha_{QED}=137.036$.

It is interesting to note that in 
Ref. \cite{Becirevic:2003pn} show 
another way to extract the experimental decay 
constant of the $\rho^{\pm}$ 
from semileptonic decay, 
$\tau^{\pm}\rightarrow \rho^{\pm}\nu_{\tau}$.
Using the PDG \cite{pdg.2012}, we get:
\beq
\label{frho_exp2}
f_{\rho^{\pm}}^{\mbox{exp}}
=
(213.8\pm 0.8) \MeV.
\enq

\subsection{$\psi(1S,2S)$ Sum Rule}

Using the mass of $\psi(3S)$ meson of $4.04 \GeV$, Fig.(\ref{spec}), 
we consider $\sqrt{s_0'}=4.0\GeV$, but in this case the decay constant of the
excited state is larger than the ground state decay constant,
so the sum rule fails.

The maximum value of $\sqrt{s_0'}$ is 3.9 GeV, where in this
case the decay constant of the excited state is slightly below
the decay of ground. The minimum value of $\sqrt{s_0'}$ is 3.7 GeV,
because $\sqrt{s_0'}$- m(2S) reaches the value of 100 MeV.

It is also interesting that the mass of the (1S) 
state is almost independent on the value of $\sqrt{s_0'}$ 
in stable Borel range, even 
varying 3.3 GeV to $\infty$, furthermore mass 
(2S) state increases with the increasing of 
$\sqrt{s_0'}$, but assumes a maximum value of
4.1 GeV.

\begin{figure}[!htb]
\begin{center}
\includegraphics[height=5cm]{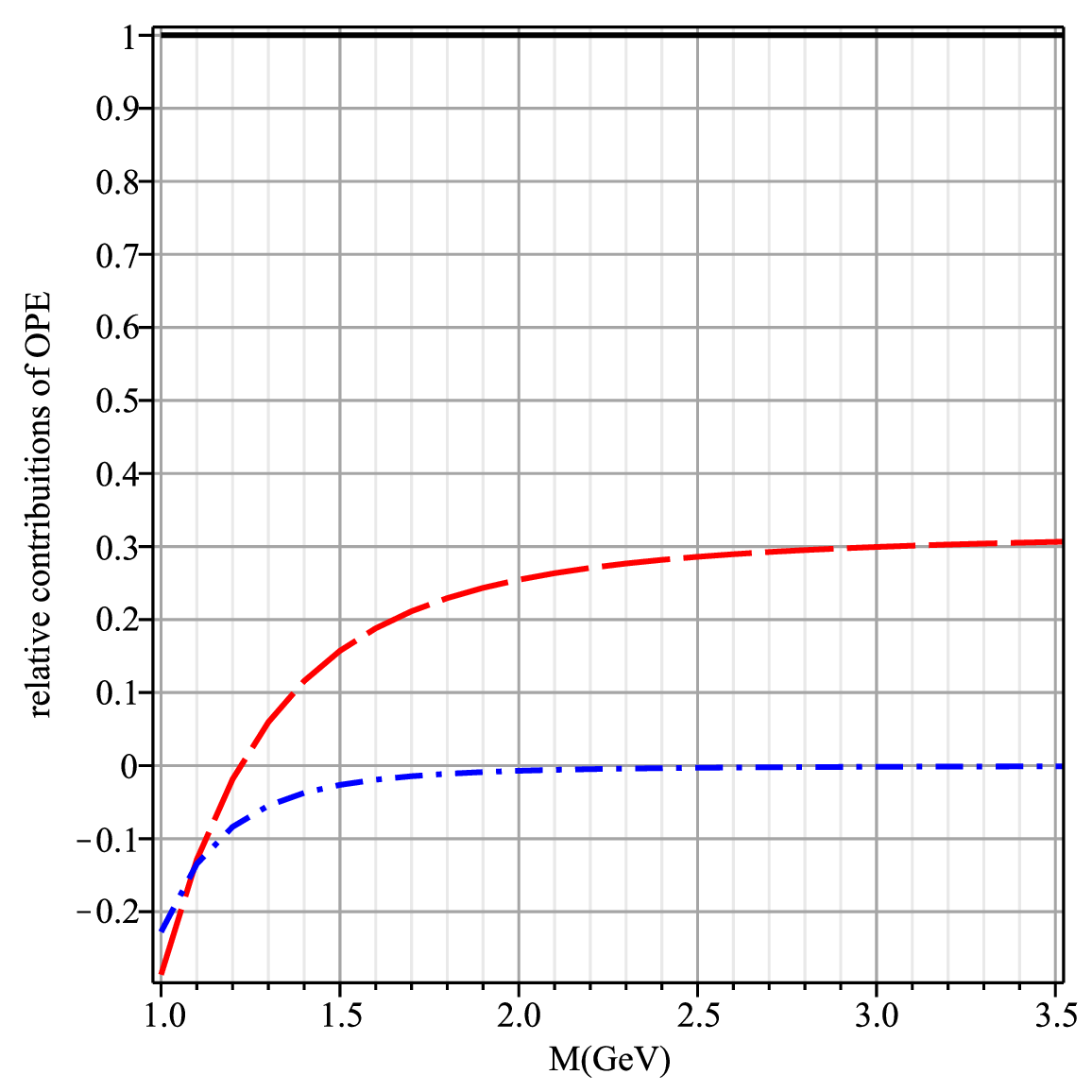}
\caption{\small The relative contributions of OPE for $\psi(1S,2S)$ 
as a function of the Borel mass for $\sqrt{s_0'}=3.8\GeV$.
The solid line for the first order perturbation term,
long-dashed line for radiative correction,
dashed-dot line for gluon condensate
.}
\label{OPEPsi}
\end{center}
\end{figure}

In Fig.(\ref{OPEPsi}), the contribution 
of the OPE terms are ordered relative 
to the first order perturbative term of 
Eq.(\ref{pert_psi}) in Eq.(\ref{corr2}). 
The solid line is the contribution 
of the first order perturbation term is adopted  as 1,
long-dash line is the radiative correction,
dash-dot line is the gluon condensate of Eq.(\ref{nonpert_psi}).
We note that the convergence of OPE is controlled 
and the contribution of the gluon condensate is
6\% of the first order perturbation term at M=1.4 GeV, 
the same order of radiative corrections.
At M=2 GeV its contribution reduces to 
only 1\% of the first order perturbation term.

\begin{figure}[!htb]
\begin{center}
\includegraphics[height=5cm]{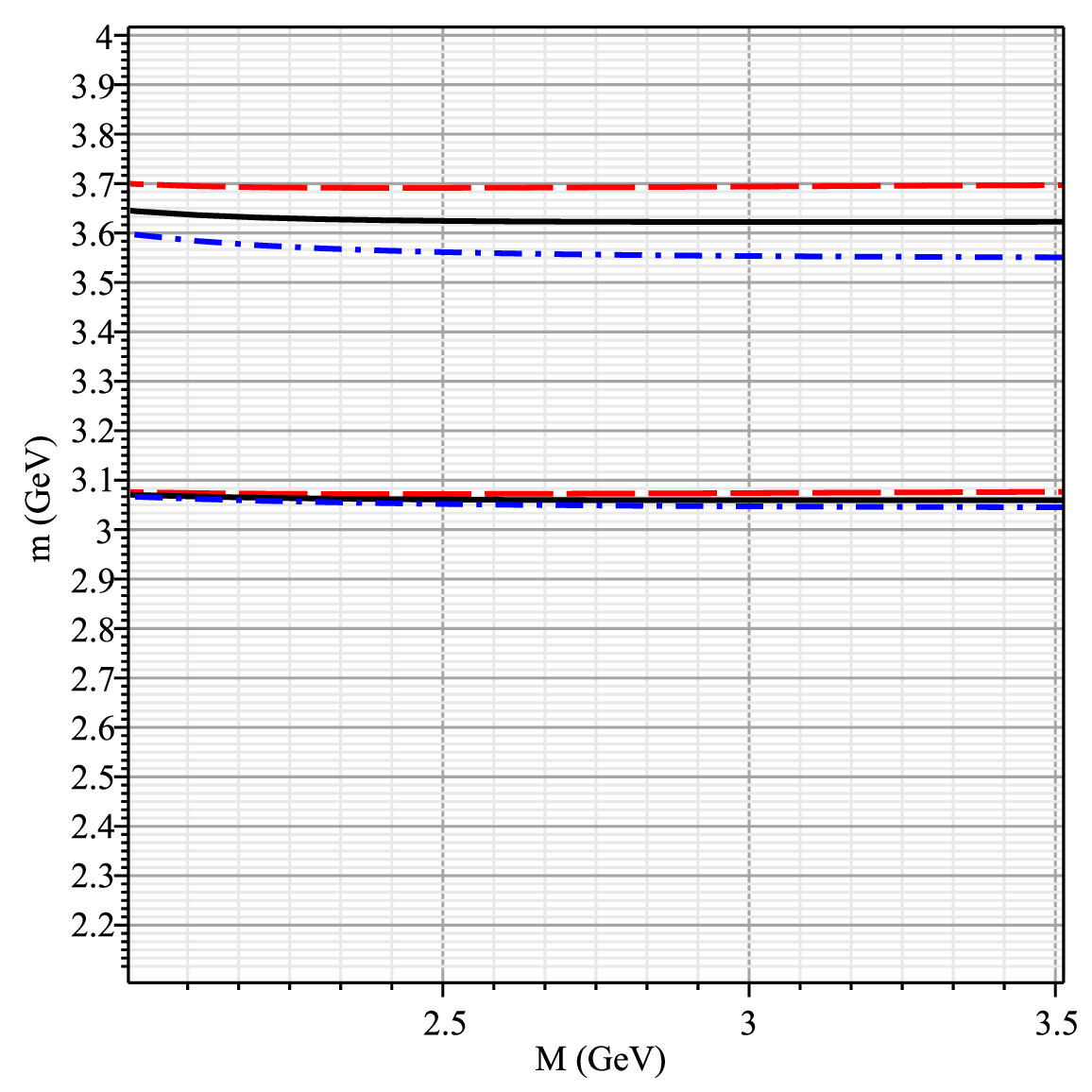}
\caption{\small The mass of $\psi(1S)$, lower lines, and $\psi(2S)$, 
upper lines, as a function of the Borel mass.
The solid line for $\sqrt{s_0'}=3.8\GeV$,
dashed-dot line for $\sqrt{s_0'}=3.7 \GeV$ and
long-dashed line for $\sqrt{s_0'}=3.9 \GeV$.}
\label{massPsi}
\end{center}
\end{figure}

We study the behavior of the mass of meson $\psi(2S)$ and $\psi$ as a function of Borel mass for three values $\sqrt{s_0'}$. We have in Fig.(\ref{massPsi}), 
solid line for $\sqrt{s_0'}=3.8 \GeV$, 
dashed-dot line for $\sqrt{s_0'}=3.7 \GeV$  and 
long-dashed line for $\sqrt{s_0'}=3.9 \GeV$. 

We can see in Fig.(\ref{massPsi}) that all masses 
are stable at $M>2$GeV and the solid line is for
the $\psi(1S)$ mass of 3.07 GeV and $\psi(2S)$ 
mass of 3.64 GeV.
%

\begin{figure}[!h]
\begin{center}
\includegraphics[height=5cm]{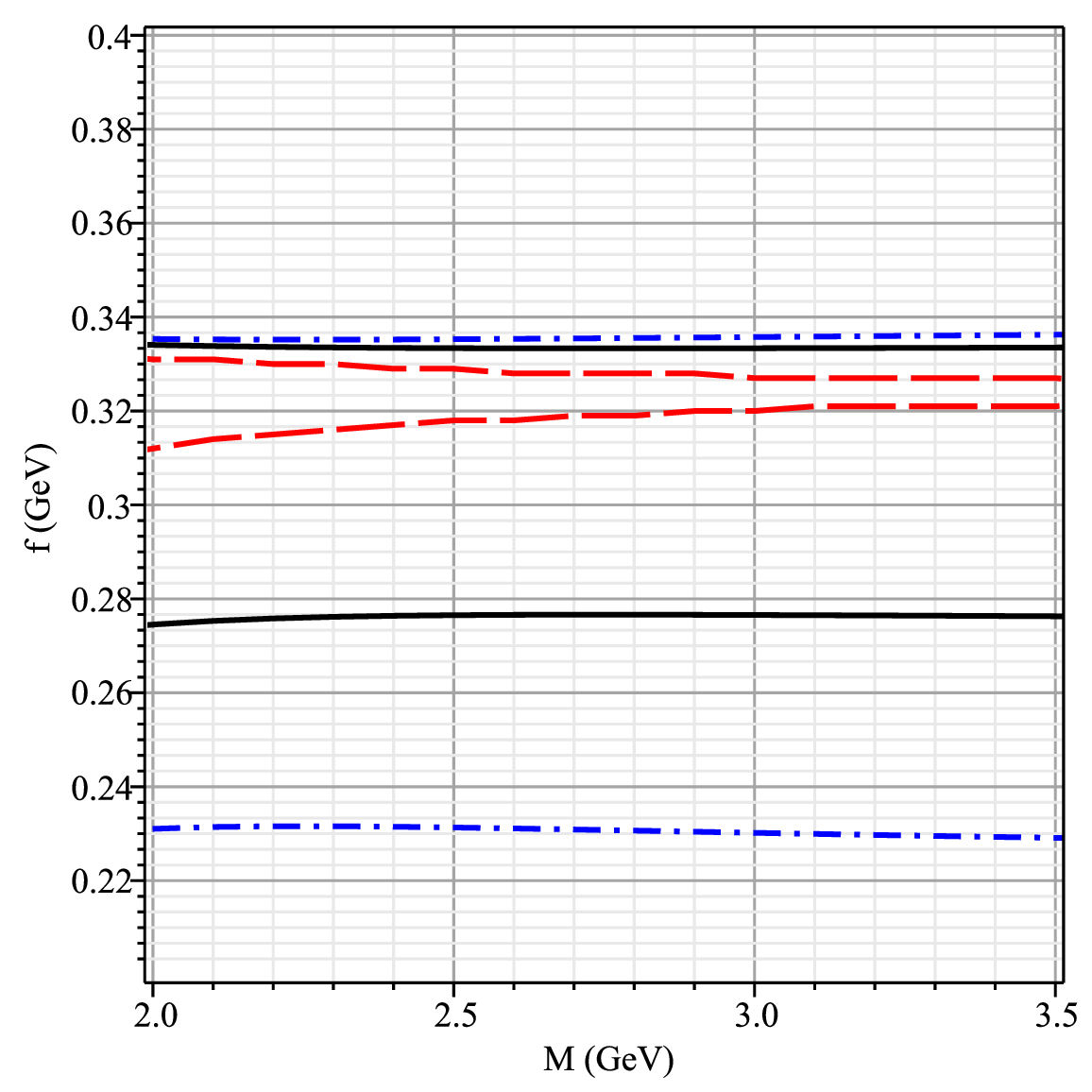}
\caption{\small The decay constant of the $J/\psi$ meson.
The solid line for $\sqrt{s_0'}=3.8\GeV$,
dashed-dot line for $\sqrt{s_0'}=3.7 \GeV$ and
long-dashed line for $\sqrt{s_0'}=3.9 \GeV$.}
\label{decayPsi}
\end{center}
\end{figure}

For the calculation of the decay constant, we use the experimental values $m_{1}=3.096 \GeV$ and $m_{2}=3.686 \GeV$. In Fig.(\ref{decayPsi}) we can see that the decay constants are stable $M>2$ GeV. 
Considering uncertainty with respect to $\sqrt{s_0'}$ parameter at $M=2$ GeV, we get:

\beq
\label{fpsi2S}
f_{\psi(2S)}
=
(272 \pm 40) \MeV,
\enq
and for $J/\psi$ meson decay constant, we get: 
$f_{\psi}=(334 \pm 1) \MeV$.  

The result for the decay constant of $\psi(2S)$ is in agreement
with the experimental value of $\psi(2S)$ of $(294 \pm 5)$ MeV \cite{pdg.2012}
obtained from $V^0\rightarrow e^{+}e^{-}$ 
decay width considering $1/\alpha_{QED}=137.036$.
For $J/\psi$, the decay constant
is 82 MeV lower than the experimental value of $J/\psi$ of
$(416\pm 5)$ MeV.
 
\subsection{$\Upsilon(1S,2S)$ Sum Rule}
 
Using the mass of $\Upsilon(3S)$ meson of $10.35 \GeV$, Fig.(\ref{spec}), we consider $\sqrt{s_0'}=10.30\GeV$.

\begin{figure}[!htb]
\begin{center}
\includegraphics[height=5cm]{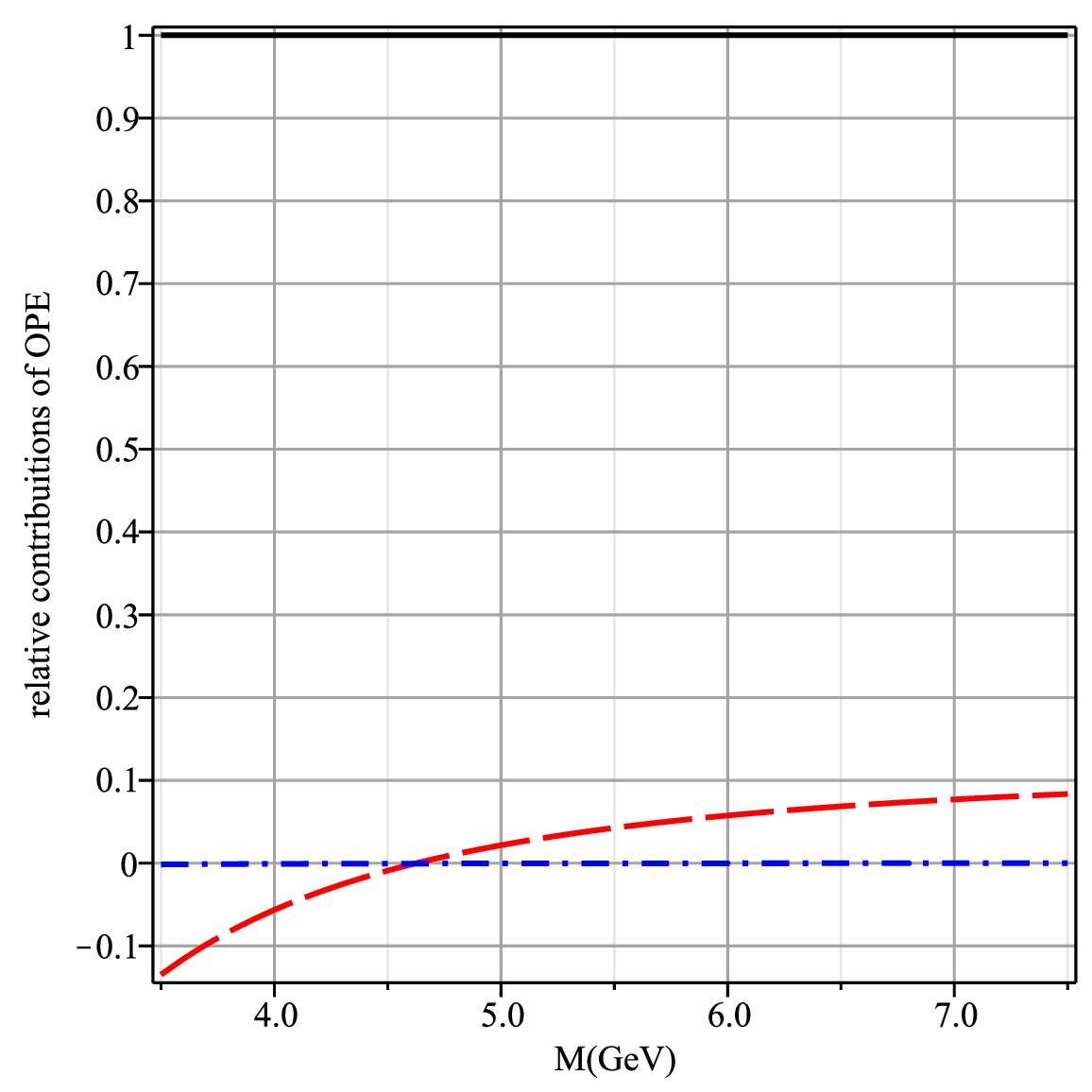}
\caption{\small The relative contributions of OPE for $\Upsilon(1S,2S)$ as a function of the Borel mass for $\sqrt{s_0'}=10.30\GeV$. 
The solid line for the first order perturbation term,
long-dashed line for radiative correction,
dashed-dot line for gluon condensate
.}
\label{OPEUpsi}
\end{center}
\end{figure}

In Fig.(\ref{OPEUpsi}), the contribution 
of the OPE terms are ordered relative 
to the first order perturbative term of 
Eq.(\ref{pert_psi}) in Eq.(\ref{corr2}). 
The solid line is the contribution 
of the first order perturbation term is adopted as 1,
long-dashed line is the radiative correction,
dashed-dot line is the gluon condensate of Eq.(\ref{nonpert_psi}).
We note that the convergence of OPE is controlled 
and the contribution of the gluon condensate is
only 0.05\% of the first order perturbation term at M=5 GeV
and 0.01\% at M=7 GeV.

\begin{figure}[!htb]
\begin{center}
\includegraphics[height=5cm]{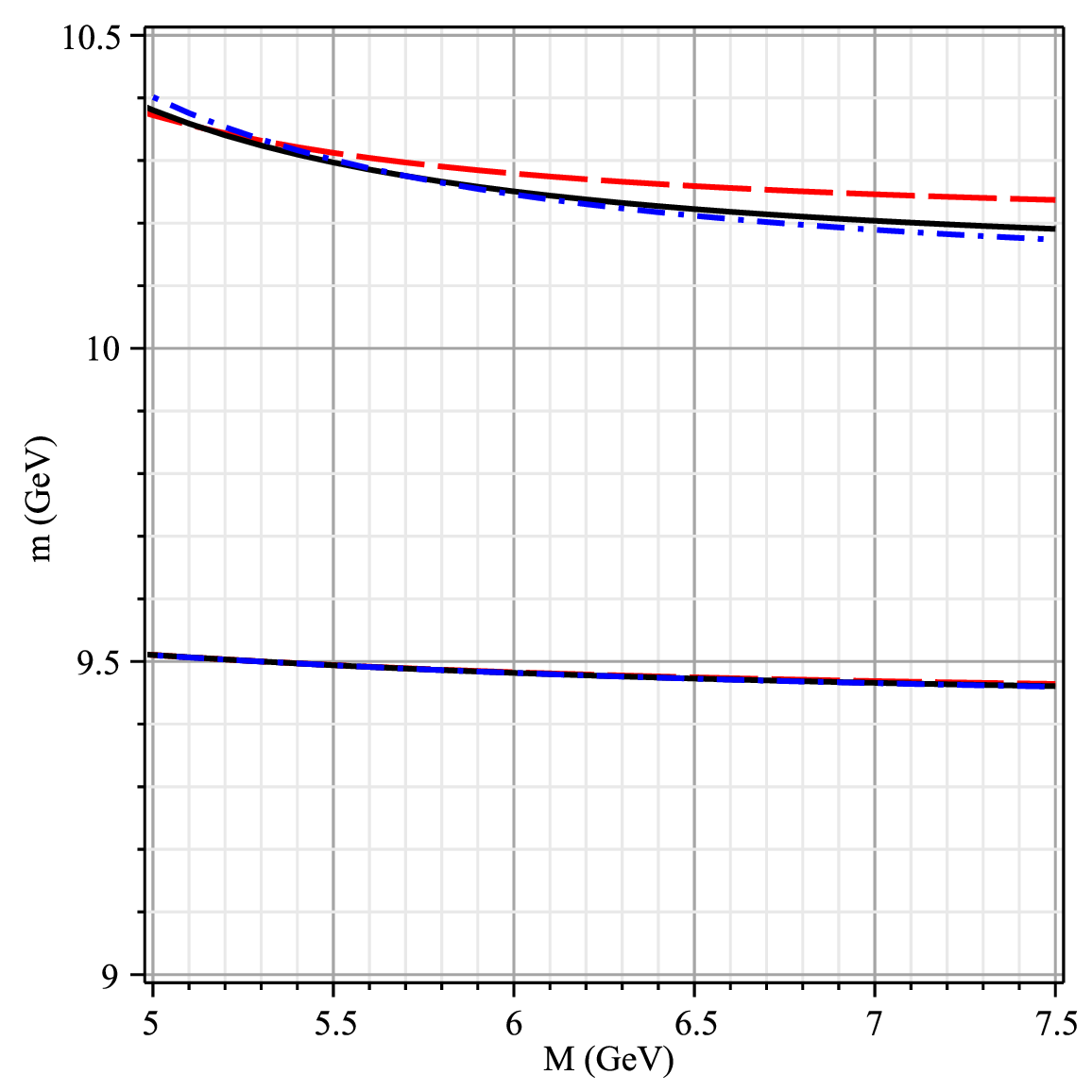}
\caption{\small The mass of $\Upsilon(1S)$, lower lines, and 
$\Upsilon(2S)$, upper lines, as a function of the Borel mass.
The solid line for $\sqrt{s_0'}=10.30\GeV$,
dashed-dot line for $\sqrt{s_0'}=10.25 \GeV$ and
long-dashed line for $\sqrt{s_0'}=10.35 \GeV$.}
\label{massUPsi}
\end{center}
\end{figure}

We study the behavior of the mass of $\Upsilon(2S)$ 
and $\Upsilon$ as a function of Borel mass for 
three values of $\sqrt{s_0'}$. 
We can see in Fig.(\ref{massUPsi}), 
solid line for $\sqrt{s_0'}=10.30 \GeV$, 
dashed-dot line for $\sqrt{s_0'}=10.25 \GeV$  
and longer dashed line for $\sqrt{s_0'}=10.4 \GeV$.

We can see in Fig.(\ref{massUPsi}) that all masses are stable at $M>6.5$ GeV.
At M=6.5 GeV, the mass obtained for $\Upsilon(1S)$ is 9.46 GeV 
and $\Upsilon(2S)$ is 200 MeV above of the experimental value.
 
\begin{figure}[!h]
\begin{center}
\includegraphics[height=5cm]{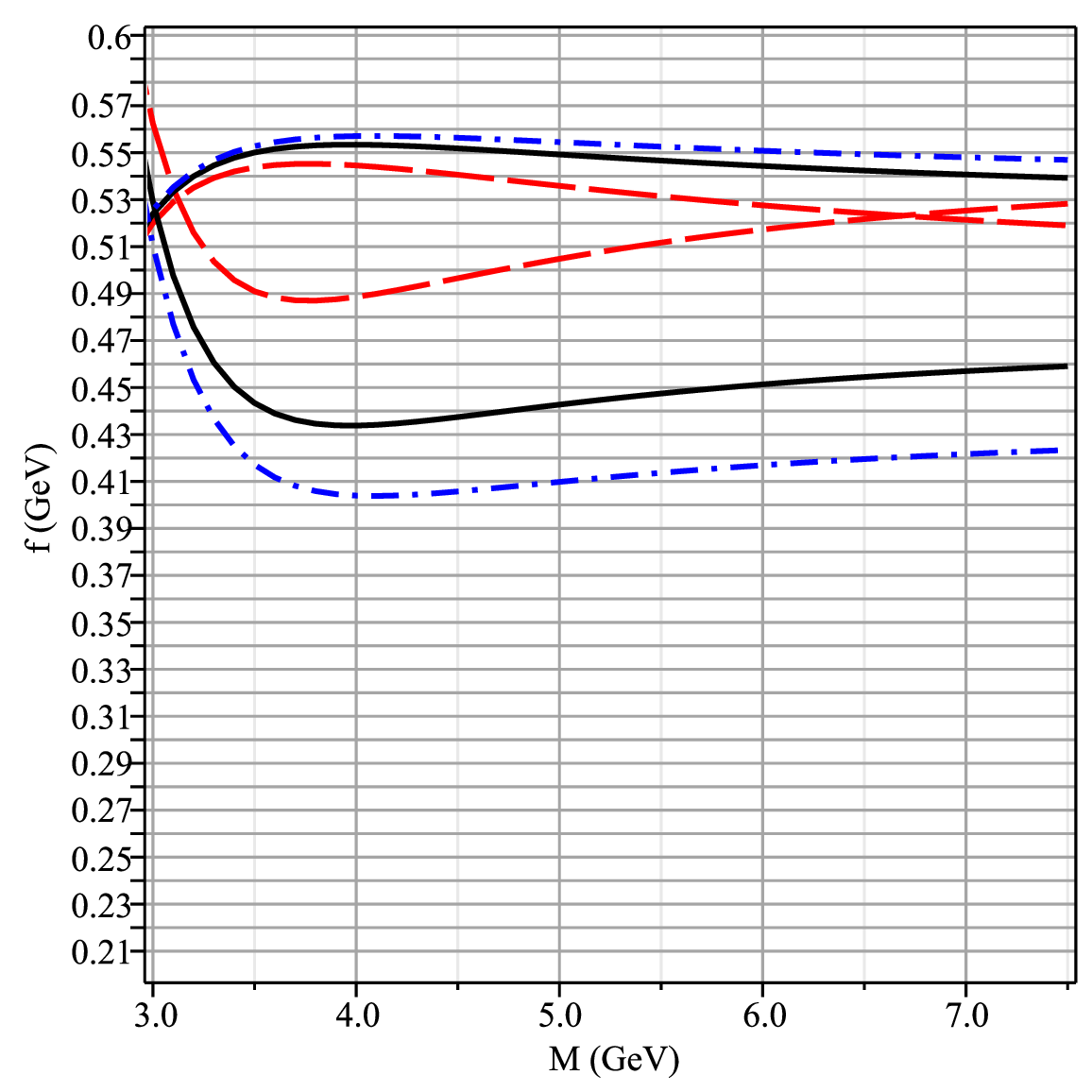}
\caption{\small 
The decay constant of $\Upsilon(2S)$, lower lines, and 
$\Upsilon(1S)$, upper lines, as a function of the Borel mass.
The solid line for $\sqrt{s_0'}=10.30\GeV$,
dashed-dot line for $\sqrt{s_0'}=10.25 \GeV$ and
long-dashed line for $\sqrt{s_0'}=10.35 \GeV$.
}
\label{decayUPsi}
\end{center}
\end{figure}

For the calculation of the decay constant, we use the experimental values $m_{1}=9.46 \GeV$, $m_{2}=10.02 \GeV$. In  Fig.(\ref{decayUPsi}) 
we see that the values for the decay constant have good stability for a Borel mass above 6 GeV. 

Considering uncertainty with respect to
$\sqrt{s_0'}$ parameter at $M=6$ GeV, 
we get:

\beq
f_{\Upsilon(2S)}=(467 \pm 51) \MeV.
\enq

For the $\Upsilon$ meson decay constant, we get: 
$f_{\Upsilon}=(540 \pm 12) \MeV$. 

\subsection{$\psi_{t}(1S,2S)$ toponium meson Sum Rule}

In this case, we show how to use our 
method to predict particles 
not yet discovered 
as $\psi_{t}(2S,1S)$ 
compound of top quark. 
One problem of this sum rule is that correction 
O($\alpha_s^2$) is important, making this sum rule 
less reliable than the $\Upsilon(1S,2S)$. 
There are papers that are against the existence 
of the toponium \cite{Fabiano:1993vx} and 
papers in favor Ref.\cite{Kiyo:2002rr,Goncharov:2008fh}
where they have predicted a mass of $\psi_{t}(1S)$ 
and $\eta_{t}(1S)$ with a mass of 347.4 GeV.

In this case $\sqrt{s_0'}$ 
is chosen to satisfy 
ordering of decay constants and 
the condition $\sqrt{s_0'}-m(2S)$ 
is of about 100 MeV.
However,  
we prefer to relax 
this condition to $\sqrt{s_0'}-m(2S)= 1\GeV$,
due the variation of m(2S) as 
Borel mass in the scale of 1 GeV.
We also consider the value of 
$m_{t}^{pole}=172$ \cite{pdg.2012},
$\alpha_s(m_t)=0.1$, where this value 
is close to the results of the Refs.\cite{KOKKAS:2014fla,Abdallah:2004xe} 
and the maximum value of the gluon condensate.

Initially, we attempt a value to $\sqrt{s_0'}$
as is shown in the first column of Tab.\ref{psit},
where the Borel window 
is limited between $M_0$ and $M_f$, where the pole contribution is above
40\% of total correlator and OPE convergence 
is controlled. 
The masses are calculated in 
distinct Borel windows. The decay constants 
are calculated at the midpoint 
$M_m=(M_0+ M_f)/2$. 
In the first attempt, we can see that the 
$\sqrt{s_0'}$ value led to a violation in 
the ordering of the decay constants, 
which leads us in the next attempt to use values of
$\sqrt{s_0'}$ smaller than 1 TeV. 
Only in the third attempt, 
was obtained the ordering. 
Now we improve the gap between $\sqrt{s_0'}$
and m(2S) with a value of about 
1 GeV, that is obtained in the final attempt. 

Thus, we get the following results for the masses of 
$\psi_t(1S,2S)$ of $m(1S)=357\GeV$ and $m(2S)=374\GeV$.
 
\begin{table}[!h]
\caption{
Sum rule of $\psi_t(1S,2S)$ for $\alpha_s(m_t)=0.1$ 
and $m_t(m_t)=164.7\GeV$. All quantities are given in 
GeV.
}
\label{psit}
\begin{tabular}{c|ccccc}
\hline
      & attempt 1              & attempt 2 
      & attempt 3              & attempt 4\\
\\   
\hline
\hline                                  
$\sqrt{s_0'}$ &1000    & 450   &376   &375  \\
$M_0$         &2000    & 200   &100   &100    \\
$M_f$         &10000   & 1000  &300   &300  \\
$m(1S)(M_0)$  & 540    & 364   &357   &357 \\
$m(2S)(M_f)$  & 903    & 430   &374   &374  \\
$f(1S)(M_m)$  & 103    & 27    &18.9  &18.7  \\
$f(2S)(M_m)$  & 109    & 32    &7.6   &7.1   \\
\hline
\end{tabular}
\end{table}

Finally, we collect all the results from the decay constant have obtained in this section in Tab.(\ref{all}).  
In the column ``this work'' refers to the extraction of 
decay constants on the same Borel window. 
The ``column experiment'' refers to the average 
values of PDG \cite{pdg.2012}, to the process $V^{0}\rightarrow e^{+}e^{-}$, considering $1/\alpha_{QED}=137.036$.

\begin{table}[!h]
\caption{Decay constants of the 2S states and 1S states in MeV. The ``column experiment'' refers to the average 
values of PDG \cite{pdg.2012}, to the process 
$V^{0}\rightarrow e^{+}e^{-}$, considering $1/\alpha_{QED}=137.036$.
In calculating the decay constant of $\rho(2S)$ 
in Ref.\cite{Yamazaki:2001er} 
we use the mass of $\rho(2S)$ of 1540 MeV
which is the average value found by them.
}
\label{all}
\begin{tabular}{c|ccccccc}
\hline
      & This work              & Ref.\cite{Arndt:1999wx} & Ref.\cite{Qin:2011xq}  
      & Ref.\cite{Peng:2012tr} & lattice                 & lattice     
      & experiment \\
      &                        &                         & ($\omega=0.5\GeV$) 
      && Ref.\cite{Yamazaki:2001er,Dudek:2006ej}&Ref.\cite{Jansen:2009hr,Becirevic:2013bsa} & Ref.\cite{pdg.2012}\\   
\hline
\hline
$\rho$                   &$203\pm 2$  &216.37  &268 &-    &$225\pm 9$   &$239\pm 18$             &$220.5 \pm 1$\\
$\rho(2S)$               &$186 \pm 14$&128     &155 &-    &$185 \pm 78$ &-             &-\\
$J/\psi$                 &$334\pm 1$  &-       &-   &-    &$399\pm 4$   &$418 \pm 13$ &$416.3 \pm 6.0$\\
$\psi(2S)$               &$272 \pm 40$&-       &-   &371  &$143\pm 81$  &-            &$294.5 \pm 4.5$\\
$\Upsilon$               &$540\pm 12$ &-        &-   &546.6&-            &-            &$715 \pm 5$\\
$\Upsilon(2S)$           &$467 \pm 51$&-       &-   &583.2&-            &-            &$497.5 \pm 4.5$\\
\hline
\end{tabular}
\end{table}

\section{Conclusions}

In this work we have presented a method to QCD sum rule with double pole 
which is basically a fit with two exponentials of the correlation function, where we 
can extract the masses and decay constants of mesons as a function of the Borel mass. 
We study the mesons: $\rho(1S,2S)$, $\psi(1S,2S)$ and $\Upsilon(1S,2S)$, where we 
know their masses and decay constants from the experimental data, 
except the $\rho(2S)$ decay constant.
We also study the hypothetica meson called toponium as an example 
how to use our method to predict new hadrons.
 
Using the experimental values for the meson masses, 
the decay constants have a good stability as Borel mass
and we have shown a prediction for the $\rho(2S)$ decay constant of 
$f_{\rho(2S)}=(186 \pm 14)$ MeV.

In addition, the decay constants of
$\psi(2S)$ and $\Upsilon(2S)$ have value lower than the experimental
values.

We finish with an application of this method 
to study the hypothetical particle called toponium. 
In this case, we start with an initial tentative value 
for the continuum threshold using a very high initial value of 1 TeV 
and we note that the ordering of the decay constants is violated, 
which led us naturally to reduce the continuum threshold up to the minimum
value of m(2S)+ 1 GeV. We use the lowest value of the continuum
threshold to get the toponiuns masses $\psi_t(1S,2S)$ of
$m(1S)=357\GeV$ and $m(2S)=374\GeV$.

\section{Acknowledgements}
{We would like to thank Prof. Francisco de Assis
de Brito for fruitful discussions. 
This work has been partially supported by CAPES.}


\end{document}